# A New Method of Capturing and Transporting Asteroids to a Lunar Orbit

**Charles Domercant[1] and Aiden Gonzalez[1]**

**31 August 2014**

[1]*Department of Aerospace Systems Design Laboratory, Georgia Institute of Technology, Atlanta, GA, 30332*

Abstract

Through multiple, small unmanned probes we can more efficiently capture an asteroid and return it to lunar orbit. The Asteroid Return and Exploration System (ARES) aims to make use of previously tested technology with the application of a new idea to satisfy a goal in the retrieval and the subsequent exploration, classification and data acquisition of an asteroid that is approximately 7 meters in diameter and 350,000 kilograms in mass. Research is conducted according to mathematical analyses of the given data and extrapolated variables.

NASA's Asteroid Recovery Mission (ARM) focuses on capturing and retrieving a near-earth asteroid. Asteroids are an integral part of our universe and their origins vary widely. They contain, in some instances, materials older than our solar system. The interest in scientific discovery surrounding a captured asteroid is tremendous. Amino acids, the basic building blocks of life, were discovered on a meteorite. If those amino acids survived entry into our atmosphere and interaction with our ecosystem and its challenges, one may extrapolate that an uncontaminated sample in space may house new forms of life. In addition to the benefits of scientific discovery, however, we may also look favorably on resource acquisition as a reason for capturing and returning an asteroid. Between 30-50% of asteroids are thought to contain water; some asteroids are theorized to contain more than 80 times what earth's largest supertanker could carry. The Asteroid Return and Exploration System (ARES) presents a new and viable method of capturing and returning a near-earth asteroid to lunar orbit.

While the capture and return of an entire asteroid has never before been attempted, there are some mission precedents in travel to and reconnaissance of asteroids. Most notably, the JAXA (Japanese Aerospace Exploration Agency) probe Hayabusa has landed on and recovered samples of the asteroid Itokawa. Further, the NASA probe Dawn has travelled to and conducted extensive studies on the minor planets Ceres and Vesta. Between probe encounters and telemetry we have managed to ascertain numerous useful scientific data, including values of albedo and their corresponding spectral type, the density and masses of asteroids, as well as their gravitational force. Though we have managed to glean an amazing amount of information from asteroids, we have never yet set foot, as the human race, on one's surface, nor have we conducted sensitive scientific tests which may only be completed by humans. Perhaps the biggest detriment to humanity, though, is that we have not yet perfected deep space travel to such an extent that we may bring back an entire asteroid.

ARES itself is a two part solution to the question of capturing an asteroid, however, several steps must be taken in preparation. First, a large space launch system (SLS) will achieve a low-earth orbit of 160 kilometers above sea level. It will then burn at the periapsis of its orbit to take advantage of the Oberth Effect and carry ARES into a transfer maneuver before jettisoning the craft. ARES itself is an unmanned craft containing several smaller probes which will travel to and make physical contact with the asteroid, securing themselves using physical attachments such as anchors and harpoons. The probes, placed strategically about the asteroid to increase torque, will stabilize the asteroid using reaction control system (RCS) thrusters, before making the return journey to lunar orbit. The ingenuous concept of ARES is that all of the thrust for the return mission is not relegated to reliance on a single craft that would have to be uncommonly large to propel an asteroid, but on several crafts to distribute the mass among thrusting units. Finally, a manned mission in the pre-existing crew capsule OSIRIS will rendezvous with the asteroid for scientific discovery and exploration. The crew will then return to earth with the collected samples and data.

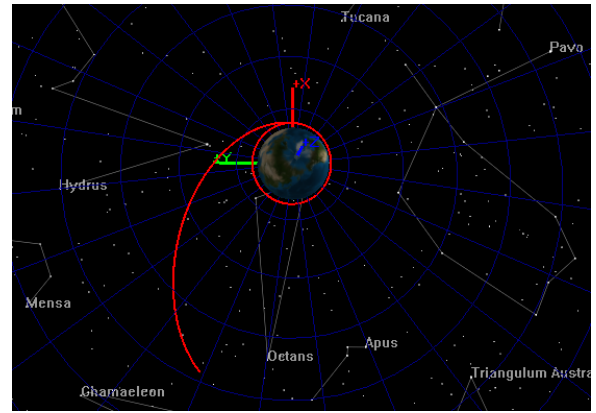

**Figure 1|** The Hohmann Transfer utilized by the SLS.

ARES proposes to be equipped with the long-term thrust solution of ion propulsion. Due to its incredibly high specific impulse ($I_{sp}$), the µ10 thruster developed by JAXA was chosen to model all mathematical experiments and scenarios. The μ10 ion thruster will use Xenon as a fuel source due to its relative abundance and affordability. The μ10 thruster has been recorded at an $I_{sp}$ as high as 3,500 seconds. The μ10 also has the benefit of being inexpensive in its energy consumption, drawing just 32 watts for each thruster.

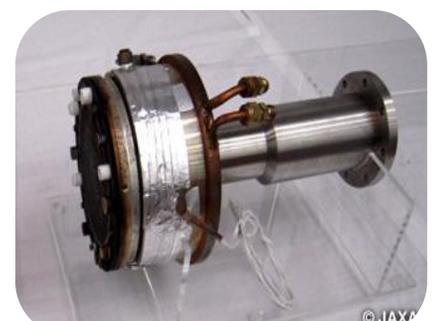

Figure 2 | The mu10 thruster from JAXA.

Because the unmanned craft's objective is not to achieve scientific discovery, but to return the asteroid for exploration and research, sparse instrumentation is needed. Only key instruments that impart communication to ground control and autonomy are required for mission success. A useful analogy to the level of autonomy planned for ARES is that of Rosetta. Due to its distance from earth, control of every aspect of Rosetta is impossible. Instead, orders are uploaded and relayed to the probe days in advance. The instrumentation of ARES should include spectrometers to determine the albedo of the asteroid, high and low gain antennae, LIDAR, gyroscopes, and a method of autonomic spatial positioning awareness, such as a sun sensor. The sparse outfit of instrumentation has, further, the added benefit of being more inexpensive in terms of mission cost.

1. **Methods**

    **Equations and Algorithms**

ΔV, literally the change in velocity, is the summation of force needed to complete a maneuver. This number is crucial to the completion of the mission. To reach the asteroid belt from a geostationary transfer maneuver, will require approximately 1,600 m/s of ΔV, to complete the return from the asteroid belt to the moon and capture its orbit will require approximately 600 m/s of ΔV, for a combined ΔV budget of 2,200 m/s or 2.2 km/s. To calculate the total amount of ΔV afforded for use by a vehicle, the equation is:

$$\Delta V = g_0 \cdot I_{sp} \cdot \ln (m_i \div m_f)$$

Using the top projected specific impulse of the µ10, 3,500 s, the preliminary calculations for the projected mass of the craft 1,630 kg, and the projected burn out mass of the craft (The mass of the structure, excluding only fuel) 200 kg the equation now becomes:

$$\Delta V = 9.8 \text{ m/s}^2 \cdot 3,500 \text{ s} \cdot \ln (1,630 \text{ kg} \div 200 \text{ kg})$$

$$\Delta V = 71,962.01 \text{ m/s}$$

This might seem very comfortably within our goal of 2,200 m/s but this models only the total possible budget of the craft, not with the added mass of the asteroid. To accurately model the trips to and fore we need two separate equations:

$$1600 \text{ m/s} = 9.8 \text{ m/s2} \cdot 3,200 \text{ s} \cdot \ln (1,630 \text{ kg} \div m_f)$$

Solving for the final mass, we get the number of 1,555.7 kg. When we subtract that from our initial mass, 1,630 – 1,555.7, we find that we have used just 74 kg of fuel. Our final mass, 1,460 kg becomes one part of our new initial mass, combined with the mass of the asteroid (which we assume to be 350,000 kg), which we divide by 5, assuming 5 of our unmanned ARES probes, giving us our second equation:

$$\Delta V = 9.8 \text{ m/s}^2 \cdot 3{,}500\text{s} \cdot \ln((1{,}555.7 \text{ kg} + (350{,}000 \text{ kg} \div 5)) \div ((350{,}000 \text{ kg} \div 5) + 200 \text{ kg}))$$

Keeping in mind that we need 600 m/s to return to a stable lunar orbit, we get:

$$\Delta V = 656.09 \text{ m/s}$$

We meet our goal with 56.09 m/s of ΔV remaining to make correctional maneuvers.

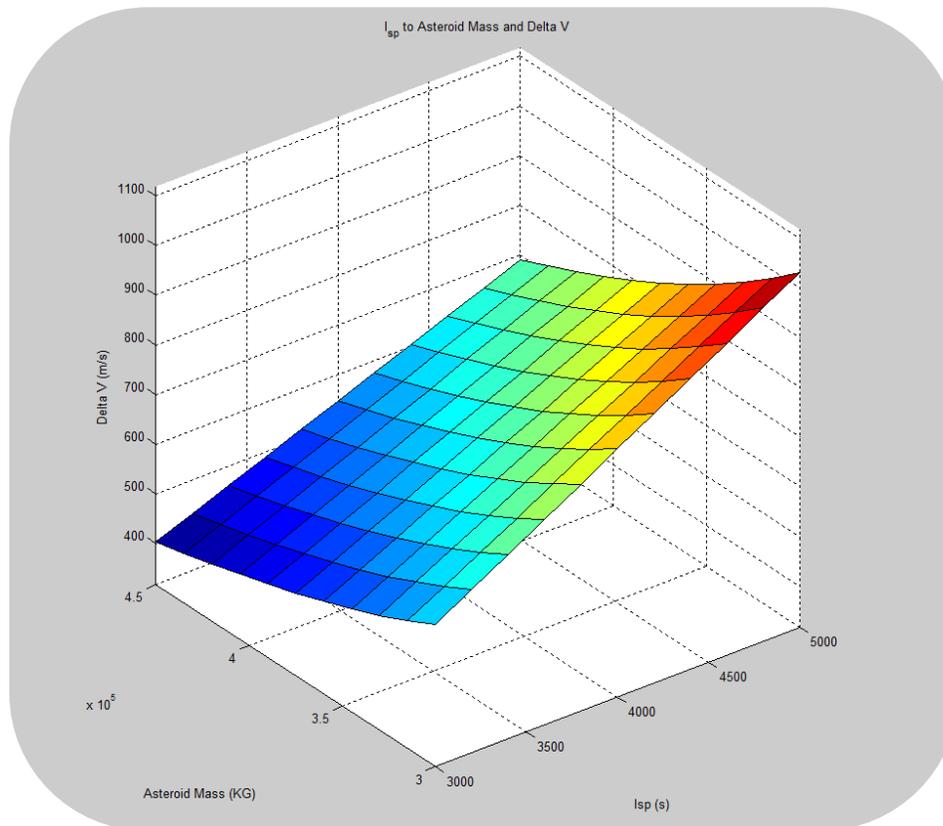

**Figure 3 |** This graph further models the ΔV available in relation to the Isp and mass of the asteroid. As the asteroid mass increases, the Isp necessary to maintain the amount of ΔV also increases proportionally.

## Conclusion and Broad Applications

Ultimately the renown and glory of being the first to return an entire asteroid should belong solely to NASA and the United States of America. It is crucial to our development to a space-faring nation that we harness objects in deep space and manipulate them to our own contrivances. The range and depth of data recovered would be extraordinary and invaluable to furthering our knowledge of the solar system and beyond.

Capturing an asteroid represents not just the retrieval and exploration of a rock, but the culmination of half a century of space travel and knowledge. It affords us the unique opportunity to acquire resources and materials from the depths of space for applications in terrestrial or even lunar infrastructure.